\documentclass[journal]{IEEEtran}
\usepackage{xcolor}
\usepackage{cite}
\usepackage{amsmath,amssymb,amsfonts}
\usepackage{algorithmic}
\usepackage{graphicx}
\usepackage{textcomp}
\def\BibTeX{{\rm B\kern-.05em{\sc i\kern-.025em b}\kern-.08em
    T\kern-.1667em\lower.7ex\hbox{E}\kern-.125emX}}
\usepackage{bm}
\newcommand{\Real}{\mathbb{R}}
\usepackage{nicefrac} %

\usepackage{amsthm}
\theoremstyle{plain}
\newtheorem{theorem}{Theorem}
\newtheorem{definition}[theorem]{Definition}

\usepackage{xspace}
\makeatletter
\DeclareRobustCommand\onedot{\futurelet\@let@token\@onedot}
\def\@onedot{\ifx\@let@token.\else.\null\fi\xspace}

\def\eg{\emph{e.g}\onedot} 
\def\ie{\emph{i.e}\onedot} 
\def\cf{\emph{c.f}\onedot} 
 
\def\wrt{w.r.t\onedot} 

\makeatother

\usepackage{svg}

\usepackage[normalem]{ulem}
\usepackage{siunitx,booktabs}
\usepackage{multirow}

\makeatletter
\let\NAT@parse\undefined
\makeatother
\definecolor{refblue}{rgb}{0,0.541,0.855}
\usepackage{hyperref}
  \hypersetup{
    colorlinks=true,
    linkcolor=refblue,
    citecolor=refblue,
    filecolor=refblue,
    urlcolor=refblue,
}
\usepackage[capitalize]{cleveref}
\crefname{section}{Sec.}{Secs.}
\Crefname{section}{Section}{Sections}
\Crefname{table}{Table}{Tables}
\crefname{table}{Tab.}{Tabs.}
\crefname{appendix}{Suppl.}{Suppl.}
\Crefname{appendix}{Supplementary}{Supplementary}

\begin{document}
\title{Benchmarking Deep Learning-Based Low-Dose CT Image Denoising Algorithms\vspace{0.75cm}}
\author{Elias Eulig, Björn Ommer, Marc Kachelrieß\vspace{0.75cm}
\thanks{This is a preprint of \url{http://doi.org/10.1002/mp.17379}.}
\thanks{Elias Eulig and Marc Kachelrieß are with the German Cancer Research Center (DKFZ), Heidelberg, Germany and Heidelberg University, Germany.}
\thanks{Björn Ommer is with the Ludwig Maximilian University Munich, Germany.}
\thanks{This work was supported in part by the Helmholtz International Graduate School for Cancer Research, Heidelberg, Germany.}
\thanks{Correspondence to \url{elias.eulig@dkfz.de}.}}
\markboth{}{}

\maketitle

\begin{abstract}
Long lasting efforts have been made to reduce radiation dose and thus the potential radiation risk to the patient for computed tomography acquisitions without severe deterioration of image quality. To this end, numerous reconstruction and noise reduction algorithms have been developed, many of which are based on iterative reconstruction techniques, incorporating prior knowledge in the projection or image domain. Recently, deep learning-based methods became increasingly popular and a multitude of papers claim ever improving performance both quantitatively and qualitatively. In this work, we find that the lack of a common benchmark setup and flaws in the experimental setup of many publications hinder verifiability of those claims. We propose a benchmark setup to overcome those flaws and improve reproducibility and verifiability of experimental results in the field. In a comprehensive and fair evaluation of several deep learning-based low dose CT denoising algorithms, we find that most methods perform statistically similar and improvements over the past six years have been marginal at best.
\end{abstract}

\section{Introduction}
\label{sec:introduction}
\IEEEPARstart{C}{omputed} tomography (CT) is an important imaging modality, with numerous applications including biology, medicine, and nondestructive testing. However, the use of ionizing radiation remains a key concern and thus clinical CT scans must follow the ALARA (as low as reasonably achievable) principle \cite{kalra2004,brenner2007}. Therefore, reducing the dose and thus radiation risk is of utmost importance and one of the primary research areas in the field.

A straightforward approach to reduce dose is by lowering the tube current (\ie, reducing the X-Ray intensity). However, this comes at the cost of deteriorated image quality due to increased image noise and thus potentially reduced diagnostic value. To alleviate this drawback, numerous algorithms have been proposed to solve the task of low-dose CT (LDCT) denoising, \ie, reducing image noise in the reconstructed image (or volume).

\looseness-1 Iterative reconstruction (IR) techniques incorporate prior knowledge in the reconstruction process and then update the reconstructed image iteratively. The prior knowledge may model statistical properties of the noise \cite{ziegler:07}, properties of the object to be reconstructed \cite{sidky:06}, or parameters of the CT system. While IR techniques can be used to reduce numerous other artifacts compared to conventional filtered back projection (FBP), they are computationally expensive, which limits their clinical applicability. On the other hand, filtering techniques to reduce noise are fast and easy to implement into various reconstruction frameworks. The filtering may either be performed in projection domain, image domain, or both, and using a wide range of algorithms \cite{balda2012,feruglio2010,li2014,manduca2009,sukovic2000}. Recently deep-learning based filtering, particularly in the image domain, became increasingly popular \cite{chen2017,chen2017c,chen2017a,fan2020,heinrich2018,huang2022a,kang2017,ramanathan2023,shan2019,wagner2022,yang2018,yang2023,zhang2021b}. The majority of the proposed methods learn a mapping from low-dose images to high-dose images in a supervised fashion using a deep neural network (DNN). Of the numerous proposed methods, most suggestions for improvement alter the network structure, loss function, or training strategy. Publications often claim ever improving performance which is commonly demonstrated by improved image quality metrics (\eg, peak signal-to-noise ratio, structural similarity) in experiments on simulated or clinical data.

In this work, we identify several flaws in the experimental setup of such methods which limit the verifiability of the claimed improvements. These include the lack of a common benchmark dataset, the use of inadequate metrics with little relation to diagnostic value, and unfair choice of hyperparameters for reference methods. Reproducibility and verifiability of scientific results, however, is paramount to scientific advancements of a field, and thus efforts towards fair benchmarking of existing and future algorithms is of utmost importance. To this end, we make the following contributions:
\begin{itemize}
    \setlength\itemsep{0.5em}
    \item We identify multiple flaws in the experimental setup of previously proposed methods which hinder the verifiability of their claimed improvements.
    \item We propose a benchmark setup\footnote{Code available at \url{https://github.com/eeulig/ldct-benchmark}.} for deep learning-based low-dose CT denoising methods, which aims to overcome those flaws and allows for a fair evaluation of existing algorithms and those yet to come.
    \item In a comprehensive and fair evaluation of several existing algorithms we find that there has been little progress over the past six years and many of the newer methods perform statistically similar or worse compared to older ones.
\end{itemize}

\section{Related Work}
In this section, we review existing works on deep learning-based LDCT denoising and image quality assessment of medical images.
\label{sec:related_work}
\subsection{Deep Learning-based LDCT Denoising} \label{sec:related_work_ldct}
CT image reconstruction aims at solving the linear system $\bm{R} x = p$, with $p\in \Real^{M}$ denoting the measurements in projection domain, $x\in\Real^{N}$ being the volume to be reconstructed, and $\bm{R}\in\Real^{M\times N}$ the Radon transform. LDCT generally aims at reconstructing $x$ using less dose, which can be \eg accomplished by lowering the tube current, thus increasing the noise in $p$ and $x$, or by lowering the number of measurements $M$, leading to sparse-view artifacts in $x$. Since previous studies indicate that DNN-based correction of the former can be superior, we here consider the task of LDCT denoising \cite{humphries2019}. Based on the domain ($p,x$, or both) in which they operate, deep-learning based methods for LDCT image denoising can be divided into three categories: projection-domain, image-domain and dual-domain.

Projection-domain methods aim to learn a mapping $f_{\theta}: p' \to p$ from low-dose projections $p'$ to high dose projections $p$, where $f_{\theta^*}$ is realized by a DNN, parameterized by weights $\theta$. These weights are either optimized in a supervised setting via
\begin{align}
    \theta^* = \underset{\theta}{\arg\min}~\mathbb{E}_{p',p \sim \mathcal{D}^{\text{train}}}\lVert f_{\theta}(p') - p \rVert \;, \label{eq:proj_supervised}
\end{align}
with $\lVert\cdot\rVert$ being some norm \cite{ma2021,yang2023a}, or unsupervised, exploiting structural similarities between adjacent projections \cite{zainulina2021,hong2023}. The denoised projections can then be reconstructed using either of the standard reconstruction techniques \cite{cormack1963,gordon1970,andersen1984}.

Image-domain methods aim to directly learn a mapping $g_{\phi}: x' \to x$ from low-dose images $x'$ (\ie, images reconstructed from low-dose projections $p'$ using FBP) to high-dose images $x$. Similar to \cref{eq:proj_supervised}, weights are typically optimized in a supervised setting, where the mean-squared error (MSE), or some other pixel- or feature-based loss between prediction and high-dose image $x$ is minimized \cite{chen2017,chen2017a,chen2017c,missert2018,heinrich2018,fan2020,zhang2021b,wagner2022,ramanathan2023,yang2023}, or $g$ is trained together with a discriminator in an adversarial fashion \cite{yang2018,shan2019,huang2022a}. Notable other works investigate unsupervised- or self-supervised training strategies, or leverage the intrinsic image prior of DNNs \cite{baguer2020}.

Lastly, dual-domain methods operate in both domains $x$ and $p$ simultaneously, by employing two separate networks $f$ and $g$, respectively. Networks are trained either separately using aforementioned loss functions \cite{yin2019,chao2022} or in an end-to-end fashion using a differentiable analytical reconstruction layer \cite{zhang2021c,zhou2021,zhou2022}.

In this work we focus on image-domain methods which dominate the research field. This is mainly due to the abundance of open source datasets, where paired high- and low-dose images are readily available \cite{mccollough2017,mccollough2020}. In contrast, projection data are generally proprietary and thus difficult to access \cite{divel2020}. The few datasets that provide them usually do so only for a (vendor-specific) subset of the data and handling of them can be cumbersome due to (hidden) preprocessing steps in the reconstruction pipeline of the vendor \cite{mccollough2020,horenko2022}. Many of the principles in the design of our benchmark setup, however, can be applied to the evaluation of projection-domain or dual-domain methods as well.
\subsection{Medical Image Quality Assessment}\label{sec:related_work_iqa}
Common full-reference quantitative measures for natural image quality assessment include the structural similarity index measure (SSIM) \cite{wang2004} and peak signal-to-noise ratio (PSNR). However, these metrics are usually not in agreement with human readers, which are considered the gold standard for image quality assessment of medical images \cite{verdun2015,renieblas2017,ohashi2023}. These are conducted by measuring the accuracy of multiple radiologists when performing some task (\eg, lesion detection or segmentation) using certain images. However, this metric relies, and is dependent, on the definition of a suitable task. Therefore, the subjective assessment of overall diagnostic quality by radiologists is a common alternative measure \cite{mason2020}. Nonetheless, since conducting multiple-reader studies is time-consuming and expensive, most algorithms for enhancement of medical images are still evaluated using quantitative metrics such as SSIM or PSNR.

In \cite{mason2020,ohashi2023}, the authors find that multiple other metrics, including the visual information fidelity (VIF) \cite{sheikh2006}, have higher correlation with human reader ratings compared to SSIM and PSNR for both CT and magnetic resonance (MR) images. Furthermore, notable recent works investigate the use of radiomic features to provide a clinically meaningful measure for the quality of medical images without the drawbacks of human reader studies \cite{pan2021,wei2021,patwari2023}.

\section{Flaws of current evaluation protocols}
\label{sec:flaws}
\begin{figure}[t]
    \centering
    \includegraphics[width=1\columnwidth]{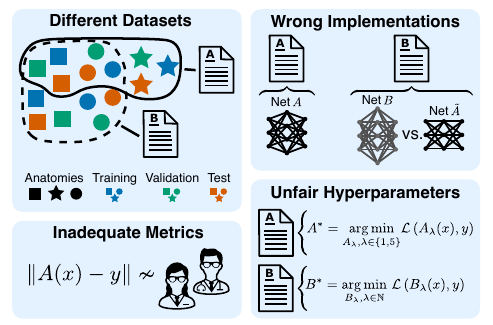}
    \caption{Overview of flaws in the experimental setup of many deep learning-based LDCT denoising methods, that limit their verifiability.}
    \label{fig:introfig}
    \vskip-\baselineskip
\end{figure}

In this section we will outline the main problems with current evaluation protocols for deep learning-based image-domain LDCT denoising (see \cref{fig:introfig} for an overview).

\subsection{Different datasets}\label{sec:flaws_datasets}
Unlike in many other disciplines of computer vision, particularly image denoising of natural images \cite{franzen1999,martin2001,zhang2011a,huang2015,zhang2017a}, there exist no consensus regarding benchmark datasets for LDCT denoising. While most methods are trained and evaluated on the dataset provided as part of the \textit{2016 NIHAAPM-Mayo Clinic LDCT Grand Challenge} \cite{mccollough2017} or the subsequently released (significantly larger both in number of images and anatomical sites) \textit{LDCT and Projection data} \cite{mccollough2020}, authors of each method employ their own training, validation, and test split. Therefore, reported metrics across publications are not comparable. This is further exacerbated by the fact that performance of individual methods differs significantly between different anatomical sites, and images (\ie, axial slices), as shown by our experiments.
\subsection{Unfair choice of hyperparameters}\label{sec:flaws_hyperparameters}
Very few publications on LDCT denoising methods report the application of hyperparameter optimization \cite{bergstra2011,bergstra2012a,snoek2012} for their own or the considered comparison methods. In none of the respective publications of the algorithms considered in this study, exhaustive hyperparameter optimization is performed. The $\nicefrac{3}{8}$ algorithms that report some form of hyperparameter optimization limit it to a grid search with few points over a single parameter (learning rate) \cite{fan2020,wagner2022}, a subset of the comparison methods \cite{fan2020}, or their own method \cite{huang2022a}. Often, authors simply use the hyperparameters reported in the reference publications \cite{chen2017a,fan2020,huang2022a}. This is particularly problematic given the choice of different datasets (\cf \cref{sec:flaws_datasets}), where hyperparameters optimized by authors of method $A$ on dataset $\mathcal{D}_A$ may not be optimal for the dataset $\mathcal{D}_B$ employed by authors $B$ in their experiments.
\subsection{Missing open source implementations}\label{sec:flaws_implementations}
With many authors not providing open source implementations of their algorithms, researchers are often left to implement comparison methods themselves. This increases the chances of errors \cite{liu2021}. Additionally, changing other aspects (such as the architecture of comparison methods \cite{fan2020}) can further bias experimental results.
\subsection{Inadequate metrics}\label{sec:flaws_metrics}
Most LDCT denoising methods are evaluated using SSIM \cite{wang2004}, PSNR, or root-mean-square error (RMSE). While these are common metrics to quantify performance for natural image denoising, they are usually not in agreement with human readers for medical images (\cf \cref{sec:related_work_iqa}), making it difficult to assess the extent to which the reported improvements actually translate into clinical benefits. This could be improved by the use of quantitative measures that are more suited for medical images (\eg, VIF), or experiments using human reader studies. In the respective publications of the eight algorithms considered in this study, however, most are evaluated using SSIM, RMSE, and PSNR only. Better metrics such as VIF or reader studies are employed in three publications only. 

\section{Benchmark setup}
\label{sec:benchmark}
In the following we present a benchmark setup to overcome the flaws of current evaluation protocols, as outlined in \cref{sec:flaws} that allows for a fair and clinically meaningful evaluation of DNNs for LDCT denoising.

\subsection{Dataset}
For our benchmark setup we utilize the \textit{Low dose CT and Projection Dataset} \cite{mccollough2020}, comprising a total of 150 scans of abdomen, head, and chest, (50 scans for each exam type) at routine dose levels. For each scan, simulated low dose-reconstructions (by means of noise insertion in the projection domain) at 25\% dose for abdomen/head and 10\% dose for chest, are available. For each exam type separately, data are split in 70\%/20\%/10\% training/validation/test set and then linearly normalized to have zero mean, unit variance. During training, we employ a weighted sampling scheme such that slices from each exam type and patient are sampled with equal probability. During testing, we reduce each scan to axial regions where the brain is present (for head scans), the lung is present (for chest scans), or the lung is not present (abdomen). The code to reproduce exact dataset splits and all preprocessing is included in our benchmark suite.
\begin{table}[tb]
\centering
\caption{Hyperparameters for all deep-learning based LDCT denoising methods considered in this study. The first three parameters are optimized for all algorithms (separately).}
\label{tab:hyperparameters_priors}
\fontsize{7.9}{\baselineskip}\selectfont
\renewcommand{\arraystretch}{0.8}
\setlength{\tabcolsep}{4pt}
{\setlength\doublerulesep{0.4pt}} 
\begin{tabular}{@{}lll@{}} 
    \toprule 
     & \bfseries Parameter & \bfseries Prior \\ 
    \midrule 
    \multirow{3}{*}{\bfseries All algorithms} & Learning rate & $\log \mathcal{U}(1 \times 10^{-5}, 0.01)$ \\
& Maximum iterations & $\mathcal{U}(1 \times 10^{3}, 1 \times 10^{5})$ \\
& Mini-batch size & $\mathcal{U}(2, 128)$ \\
    \midrule\midrule 
    \bfseries CNN-10 (2017) \cite{chen2017} & Patchsize & $\mathcal{U}(32, 128)$ \\
    \midrule 
    \bfseries RED-CNN (2017) \cite{chen2017a} & Patchsize & $\mathcal{U}(32, 128)$ \\
    \midrule 
    \multirow{4}{*}{\bfseries WGAN-VGG (2017) \cite{yang2018}} & $\beta_1$ of Adam & $\mathcal{U}(0.3, 0.9)$ \\
& Loss weight: $\lambda_\text{perceptual}$ & $\mathcal{U}(0, 1)$ \\
& Critic updates & $\mathcal{U}(1, 5)$ \\
& Patchsize & $\mathcal{U}(32, 128)$ \\
    \midrule 
    \bfseries ResNet (2018) \cite{missert2018} & Patchsize & $\mathcal{U}(32, 128)$ \\
    \midrule 
    \bfseries QAE (2019) \cite{fan2020} & Patchsize & $\mathcal{U}(32, 128)$ \\
    \midrule 
    \multirow{7}{*}{\bfseries DU-GAN (2021) \cite{huang2022a}} & $\beta_1$ of Adam & $\mathcal{U}(0.3, 0.9)$ \\
& Cutmix warmup & $\mathcal{U}(0, 1 \times 10^{4})$ \\
& Loss weight: $\lambda_\text{adv}$ & $\mathcal{U}(0, 1)$ \\
& Loss weight: $\lambda_\text{CM}$ & $\mathcal{U}(0, 10)$ \\
& Loss weight: $\lambda_\text{px,grad}$ & $\mathcal{U}(0, 40)$ \\
& Critic updates & $\mathcal{U}(1, 5)$ \\
& Patchsize & $\mathcal{U}(32, 128)$ \\
    \midrule 
    \bfseries TransCT (2021) \cite{zhang2021b} & - & - \\    \midrule 
    \multirow{4}{*}{\bfseries Bilateral (2022) \cite{wagner2022}} & Learning rate for $\sigma_r$ & $\log \mathcal{U}(1 \times 10^{-5}, 0.01)$ \\
& Patchsize & $\mathcal{U}(32, 128)$ \\
& Initalization for $\sigma_r$ & $\mathcal{U}(0, 1)$ \\
& Initalization for $\sigma_{x,y}$ & $\mathcal{U}(0, 1)$ \\
    \bottomrule 
\multicolumn{3}{p{240pt}}{$\mathcal{U}$: Uniform distribution; $\log \mathcal{U}$: Log-uniform distribution}
\end{tabular} 

\end{table}
\subsection{LDCT denoising algorithms}
We consider eight DNN-based LDCT denoising algorithms proposed in the literature over the past six years. In the following we briefly describe each of the methods and refer the reader to the respective publications for more details. \textbf{CNN-10 (2017) \cite{chen2017}} is a simple three layer CNN, trained to minimize the MSE between network output and high-dose targets. \textbf{RED-CNN (2017) \cite{chen2017a}} and \textbf{ResNet (2018) \cite{missert2018}} are trained in the same fashion but employ deeper network architectures with residual connections compared to CNN-10. \textbf{WGAN-VGG (2017) \cite{yang2018}} and \textbf{DU-GAN (2021) \cite{huang2022a}} are trained in an adversarial fashion \cite{goodfellow2014,arjovsky2017}, where DU-GAN utilizes a U-net based discriminator \cite{schonfeld2020}. \textbf{QAE (2019) \cite{fan2020}} is based on RED-CNN in both network architecture and training scheme, but employs quadratic convolutions. \textbf{TransCT (2021) \cite{zhang2021b}} is based on transformer blocks and also trained with an MSE loss. \textbf{Bilateral (2022) \cite{wagner2022}} uses a trainable bilateral filter instead of a DNN, and thus substantially reduces the amount of free model parameters.
\subsection{Hyperparameter optimization}\label{sec:benchmark_hpopt}
As discussed in \cref{sec:flaws_hyperparameters}, for none of the methods a rigorous hyperparameter optimization was employed in the original publications. To ensure a fair comparison between different algorithms we optimize hyperparameters as follows. For each method we first identify hyperparameters and their suitable ranges. This includes general parameters such as learning rate, mini-batch size, patchsize and number of iterations, but also weighting factors in the loss functions (\eg, to balance adversarial and pixelwise loss in a GAN setting). Suitable ranges were determined from the respective papers (with sufficient margin) and whenever two methods had the same hyperparameter (\eg, learning rate or patchsize), we kept the prior distribution over the search space the same. All hyperparameters and their respective prior distributions are reported in \cref{tab:hyperparameters_priors}. For each method, we then performed a black box hyperparameter tuning using sequential-model based optimization (SMBO). Such an automatic approach is preferred over manual (human) optimization as it avoids any potential bias by the practitioner, thus ensuring fair comparison of different models. Furthermore, SMBO has been shown to outperform both human optimization and non-sequential optimization schemes like grid search or random search on a variety of DNN and dataset combinations \cite{bergstra2011,snoek2012}.

Let $t_\lambda:\{f_{\theta}, \mathcal{D}^{\text{train}}, \lambda\} \to \theta^*$ denote the outcome of some training run of network $f$ on training data $\mathcal{D}^{\text{train}}$ using hyperparameters $\lambda$. The aim of hyperparameter optimization is to find an optimial set of hyperparameters $\lambda^*$,\ie,
\begin{align}
    \lambda^*&= \underset{\lambda}{\arg\max}~\mathbb{E}_{x,y\sim\mathcal{D}^{\text{val}}} \left[ M\left(y, f_{t_\lambda}(x) \right)\right] \notag \\
    &= \underset{\lambda}{\arg\max}~\Psi(\lambda) \;, \label{eq:hp_opt}
\end{align}
where $M$ is some metric and $\mathcal{D}^{\text{val}}$ the validation dataset (not used during $t_\lambda$). Since evaluating $\Psi(\lambda)$ is expensive, requiring a full training run $t_\lambda$, one uses a probabilistic model $p_\Psi$, here constructed via Gaussian Processes, as a surrogate for $\Psi$. For each iteration in the optimization process, we then find the most promising next point $\lambda$, to run the costly evaluation $\Psi(\lambda)$ for, by maximizing some acquisition function. In our experiments we used the expected improvement (EI) \cite{bergstra2011} as acquisition function:
\begin{align}
    \text{EI}(\lambda, \Psi^*) = \int \max(z-\Psi^*,0) ~p_\Psi(z|\lambda)dz \;,
\end{align}
where $\Psi^*$ refers to the expectation of $M$ on the validation data for the best set of hyperparameters found so far (\ie, the one that maximizes the r.h.s.\ of \cref{eq:hp_opt} up to now). As metric $M$ that is optimized by the hyperparameter optimization, we used the SSIM for all networks. Optimizing the SSIM is favorable over other measures, since it is fast to compute, unlike \eg, VIF, and not directly involved in the training process $t_\lambda$ of any of the methods considered in this study (unlike \eg, RMSE). Further, note that for methods using a vanilla GAN loss, \eg, \cite{huang2022a}, simply minimizing the validation loss would not be suitable as it is not directly related to training progress.
For each method, we perform 50 iterations of SMBO, sufficient to ensure convergence for all algorithms, as verified by our experiments.

After an optimal set of hyperparameters $\lambda^*$ was found, we retrained a method using $\lambda^*$ ten times with different random seeds. If not stated otherwise, all reported standard deviations and significance tests (to compare two methods) are computed over those ten training runs.

\subsection{Metrics}
We evaluate all methods on the same test set comprising a total of 15 scans (5 head/chest/abdomen) using three common full-reference measures of image quality: SSIM, PSNR\footnote{\color{black} We here omit evaluation of RMSE since it is related to the PSNR via $\text{PSNR}=20\log_{10} \left(I_\text{max} / \text{RMSE}\right)$, with $I_\text{max}$ being the maximum pixel value.}, and VIF. As described in \cref{sec:flaws_metrics}, both SSIM and PSNR are common metrics to evaluate DNNs for LDCT denoising. We include VIF, since it has been shown to have higher correlation with human readers for medical images \cite{mason2020,ohashi2023}. 

Conducting human reader studies is time-consuming and expensive and would render the application of the proposed benchmark setup to future algorithms impossible. To nevertheless evaluate the algorithms in terms of clinically relevant image properties, we include an analysis of radiomic features. To this end, we compare the similarity of radiomic features extracted on the denoised images to those extracted on the high-dose image.
\begin{table*}[t]
\centering
\caption{Summary of results on the metrics SSIM, PSNR, and VIF. We highlighted a metric bold, if it is significantly better than the previously best method on that anatomy. Likewise, we highlighted a metric italics, if it is significantly worse than the previously best method on that anatomy.}
\label{tab:results_metrics}
\fontsize{7}{\baselineskip}\selectfont
\setlength{\tabcolsep}{3pt}
\robustify\bfseries
\robustify\itshape
\robustify\uline
\sisetup{uncertainty-mode=separate, round-mode=uncertainty, round-precision=1, table-align-uncertainty=true, mode=text, detect-mode=true, detect-weight=true, detect-shape=true}
\begin{tabular}{@{}lS[table-format = 1.4(1)] S[table-format = 2.2(1)] S[table-format = 1.4(1)] S[table-format = 1.4(1)] S[table-format = 2.2(1)] S[table-format = 1.3(1)] S[table-format = 1.3(1)] S[table-format = 2.1(1)] S[table-format = 1.3(1)] r@{}} 
    \toprule 
    & \multicolumn{3}{c}{{Chest (10\% dose)}} & \multicolumn{3}{c}{{Abdomen (25\% dose)}} & \multicolumn{3}{c}{{Head (25\% dose)}} & \multirow{ 2}{*}{Rank} \\
    \cmidrule(lr){2-4}\cmidrule(lr){5-7}\cmidrule(lr){8-10} 
                 & {SSIM} & {PSNR (dB)} & {VIF} & {SSIM} & {PSNR (dB)} & {VIF} & {SSIM} & {PSNR (dB)} & {VIF} \\ 
    \midrule 
    LD & 0.34      & 18.77     & 0.09      & 0.84      & 28.67     & 0.34      & 0.88      & 26.4      & 0.55 & 9\textsuperscript{ }     \\ 
  CNN-10  (2017) & \bfseries 0.5867 +- 0.0006 & \bfseries 27.7086 +- 0.0244 & \bfseries 0.1915 +- 0.0008 & \bfseries 0.8959 +- 0.001 & \bfseries 32.3924 +- 0.0979 & \bfseries 0.449 +- 0.0029 & \bfseries 0.8961 +- 0.0037 & \bfseries 28.8595 +- 0.5553 & \bfseries 0.6203 +- 0.0058 & 3\textsuperscript{ }\\ 
  RED-CNN (2017) & \bfseries 0.6086 +- 0.0018 & \bfseries 28.3578 +- 0.032 & \bfseries 0.2205 +- 0.0026 & \bfseries 0.9028 +- 0.0007 & \bfseries 33.223 +- 0.071 & \bfseries 0.4913 +- 0.0077 & \bfseries 0.904 +- 0.001 & \bfseries 30.4075 +- 0.1513 & \bfseries 0.6856 +- 0.012 & 1\textsuperscript{ }\\ 
  WGAN-VGG (2017) & \itshape 0.5086 +- 0.0304 & \itshape 25.5378 +- 0.2176 & \itshape 0.148 +- 0.0035 & \itshape 0.8821 +- 0.0023 & \itshape 30.5092 +- 0.8541 & \itshape 0.3822 +- 0.0098 & \itshape 0.8835 +- 0.0153 & \itshape 25.3598 +- 2.7384 & \itshape 0.5321 +- 0.0156 & 6\textsuperscript{\textdagger}\\ 
  ResNet (2018) & \bfseries 0.6103 +- 0.0013 & \bfseries 28.4174 +- 0.0274 & \bfseries 0.2235 +- 0.0016 & \itshape 0.9012 +- 0.0023 & \itshape 33.1527 +- 0.0784 & \itshape 0.4869 +- 0.0063 & 0.9008 +- 0.005 & \itshape 29.6392 +- 0.8303 & \itshape 0.6718 +- 0.0167 & 2\textsuperscript{ }\\ 
  QAE (2019) & \itshape 0.5844 +- 0.0025 & \itshape 27.6212 +- 0.0919 & \itshape 0.1863 +- 0.0034 & \itshape 0.8942 +- 0.0015 & \itshape 32.0231 +- 0.1717 & \itshape 0.4181 +- 0.0071 & \itshape 0.8991 +- 0.0013 & \itshape 28.5094 +- 0.261 & \itshape 0.5944 +- 0.0083 & 5\textsuperscript{ }\\ 
  DU-GAN (2021) & \itshape 0.5654 +- 0.0044 & \itshape 26.6802 +- 0.1102 & \itshape 0.1681 +- 0.0016 & \itshape 0.8935 +- 0.0017 & \itshape 32.1266 +- 0.2692 & \itshape 0.4268 +- 0.0047 & 0.9027 +- 0.0026 & \itshape 28.7625 +- 1.0176 & \itshape 0.6215 +- 0.0054 & 4\textsuperscript{ }\\ 
  TransCT (2021) & \itshape 0.5628 +- 0.0019 & \itshape 26.9851 +- 0.0547 & \itshape 0.167 +- 0.0016 & \itshape 0.8765 +- 0.0026 & \itshape 30.5286 +- 0.1749 & \itshape 0.3718 +- 0.0072 & \itshape 0.8494 +- 0.0047 & \itshape 24.6521 +- 0.3973 & \itshape 0.4387 +- 0.0139 & 6\textsuperscript{\textdagger}\\ 
  Bilateral (2022) & \itshape 0.5545 +- 0.0011 & \itshape 25.5885 +- 0.0412 & \itshape 0.1594 +- 0.0015 & \itshape 0.8591 +- 0.0025 & \itshape 27.1312 +- 0.1496 & \itshape 0.3611 +- 0.0027 & \itshape 0.8731 +- 0.002 & \itshape 26.6 +- 0.138 & \itshape 0.5004 +- 0.0043 & 8\textsuperscript{ }\\ 
\bottomrule
\end{tabular} 

\end{table*}

\begin{definition}[Radiomic feature similarity]
Let $\cos (x,y)$ be the cosine similarity between two vectors x and y:
\begin{align}
    \cos (x,y) = \frac{x\cdot y}{\lVert x \rVert \lVert y \rVert} \; .
\end{align}
Further, let $A=\{0, 1, 2, \dots,n\}$, with $n$ being the number of algorithms considered, and index $0$ being associated with the high-dose image. We denote with $R_{i,j}^{(s)}$ the radiomic feature $j\in \{1,2,\dots, J \}$ extracted on scan $s$ associated with algorithm $i\in A$. In order to get a task-agnostic metric, we assign an equal a-priori importance to each feature by normalizing
\begin{align}
    \tilde{R}_{i,j}^{(s)} = \frac{R_{i,j}^{(s)} - \underset{k\in A}{\max}{R_{k,j}^{(s)}}}{ \underset{k\in A}{\max}{R_{k,j}^{(s)}} - \underset{k\in A}{\min}{R_{k,j}^{(s)}}} \; .
\end{align}
The radiomic feature similarity $\text{RFS}_i^{(s)}$ of algorithm $i=1,...,n$ on some scan $s$ is then given as
\begin{align}
    \text{RFS}_i^{(s)} = \cos \left(r_i^{(s)}, r_0^{(s)} \right), \quad r_i^{(s)} = \left(\tilde{R}_{i,1}^{(s)}, \dots \tilde{R}_{i,J}^{(s)} \right)\;. \label{eq:rad_feat_sim}
\end{align}
\end{definition}
Radiomic features are commonly extracted on segmentations of tumors or entire organs. On the high-dose scans of the test data, we therefore segment the following organs using the TotalSegmentator \cite{wasserthal2023}: lung on chest scans, liver on abdomen scans, and brain on head scans. This segmentation mask is then used for subsequent extraction of 91 radiomic features\footnote{This includes features from the following classes (\# of features): first order statistics (18), gray level co-occurrence matrix (24), gray level run length matrix (16), gray level size zone matrix (16), neighbouring gray tone difference matrix (4), and gray level dependence matrix (13).} using PyRadiomics \cite{vangriethuysen2017}.
\subsection{LDCT-hard benchmark dataset}
In our experiments we find that the performance of all algorithms varies greatly, both between different exam types and images of the same exam type. The latter observation motivates us to derive a novel collection of test datasets, each of which being a subset of the \textit{Low dose CT and Projection Dataset} \cite{mccollough2020}. We refer to  LDCT-hard-$q$\%, as the subset containing the $q\%$ slices with lowest average SSIM across all evaluated methods. To not underrepresent anatomies for which methods achieve generally higher SSIMs (\eg, head), this subset is collected for each exam type separately.

\section{Results}
\label{sec:results}
\subsection{Hyperparameter optimization}
\begin{figure}[tb]
  \centering
  \includegraphics[width=\columnwidth]{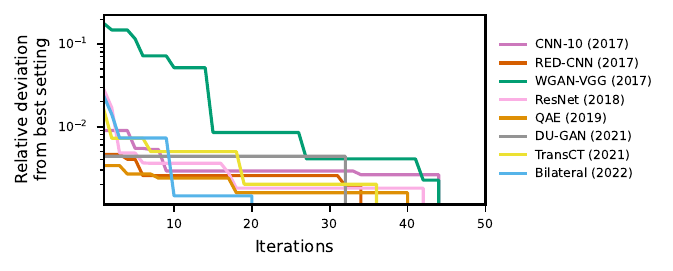}
  \vskip-\baselineskip
  \caption{Evolution of relative deviation from best setting over the 50 iterations of bayesian hyperparameter optimization. For each iteration $i$ we show the relative deviation of the best network up to $i$ from the final best configuration of hyperparameters (over all 50 iterations).}
  \label{fig:convergence_hyperparameters}
  \vskip-\baselineskip
\end{figure}
We first verify that all methods converged within the 50 iterations of Bayesian hyperparameter optimiztaion (\cref{fig:convergence_hyperparameters}). To this end, we evaluate for each method and iteration $i$ the relative deviation $\text{RelDev}_i$ from the best setting \wrt the SSIM on the validation set
\begin{align}
    \text{RelDev}_i = 1 - \frac{\max_{j\leq i} \text{SSIM}_j}{\max_j \text{SSIM}_j} \;.
\end{align}
We find that hyperparameter optimization for most of the methods converged within the first 40 iterations and none of the methods improved in the last five iterations (\cf intercept with x-axis in \cref{fig:convergence_hyperparameters}). For all methods $\text{RelDev}_{i\geq30}<0.5\%$.
\subsection{Evaluation using standard image quality metrics}
\looseness-1 We then evaluate all algorithms using the following image quality metrics: SSIM, PSNR, and VIF (\cref{tab:results_metrics}). For each method, we test if it performs significantly better or worse than the previously best method, using the nonparametric \textit{Mann-Whitney U} test \cite{mann1947} with significance level $\alpha=5\%$. While we find that ResNet significantly outperforms previous methods on the chest data, none of the newer methods consistently outperforms RED-CNN, one of the earliest deep-learning based methods for LDCT denoising (\cf \textbf{bold} numbers in \cref{tab:results_metrics}). On the contrary, for many configurations newer methods perform significantly worse than RED-CNN (\cf \textit{italic} numbers in \cref{tab:results_metrics}). In particular, we find that the two newest methods considered in this study (TransCT and Bilateral) perform significantly worse \wrt all metrics and exam types compared to RED-CNN. Remarkably, they even perform significantly worse compared to the low-dose scan on few metric and exam type combinations (\eg, TransCT on head scans for all metrics; Bilateral on abdomen scans for PSNR).

\begin{figure*}[tbp]
  \centering
  \includegraphics[width=0.95\textwidth]{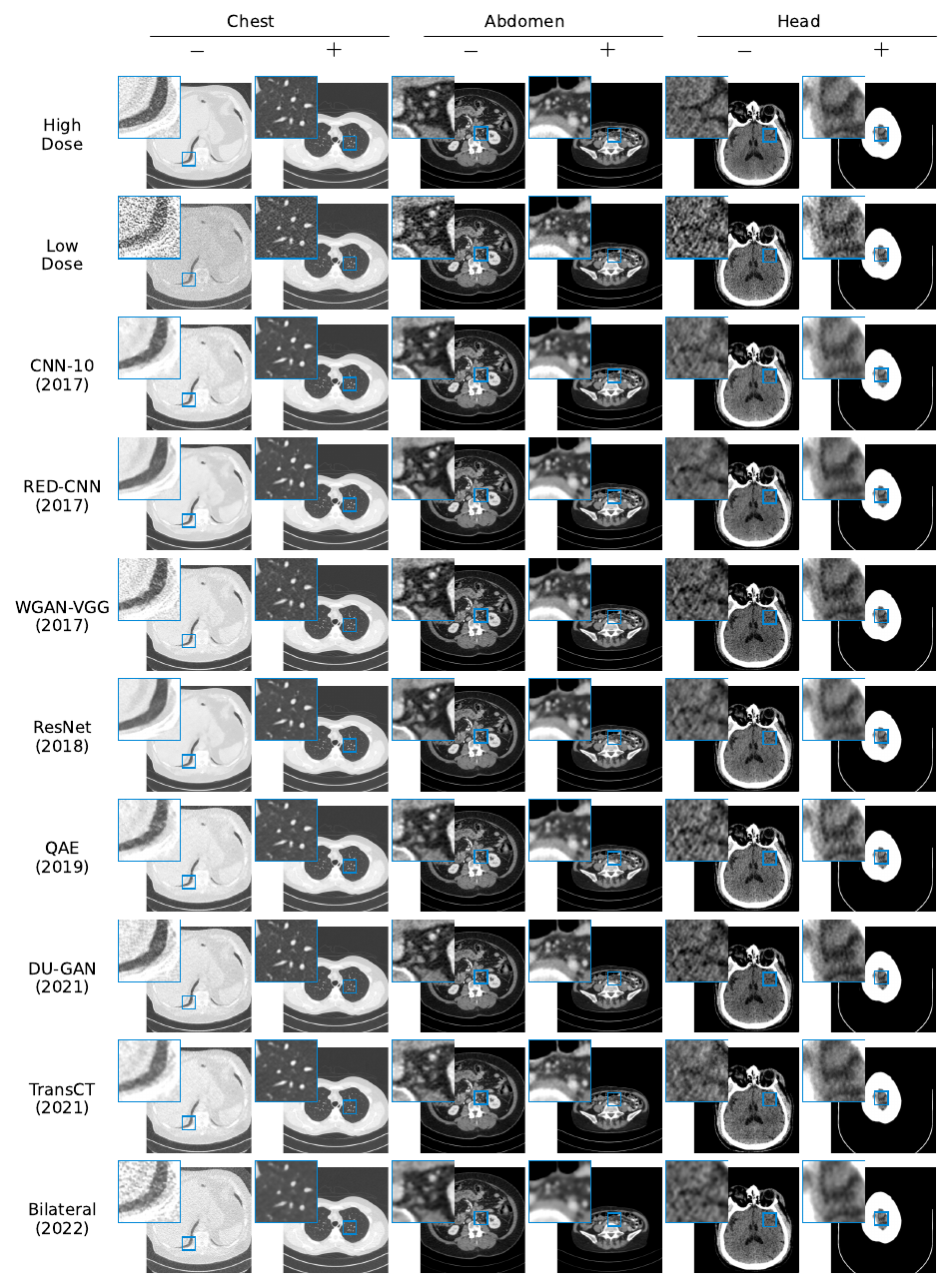}
  \caption{Best viewed zoomed in. Slices from the test dataset, for which the average SSIM over all methods is lowest (-) and highest (+). For each method, we show results for the best performing network (over the ten random trials), \ie, network having the highest SSIM on the validation data.}
  \label{fig:hardest_easiest}
\end{figure*}
\subsection{Evaluation using radiomic feature similarity}
\begin{figure}[tb]
    \centering
    \includegraphics[width=\columnwidth]{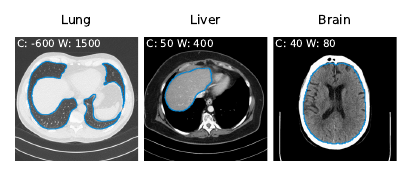}
    \vspace{-2\baselineskip}
    \caption{Contour plots of automatic segmentations for three high-dose scans of the test set of lung, liver, and brain. Radiomic features were extracted within these segmentations for low- and high-dose as well as all denoised volumes.}
    \label{fig:segmentations}
\end{figure}
\begin{table}[tb]
    \centering
    \caption{Radiomic feature similarity for three organ segmentations. Bold numbers indicate that a method is significantly better than the previously best method on that anatomy. Likewise, italics indicate that it is signficantly worse.}
    \label{tab:results_radsim}
    \fontsize{7.5}{\baselineskip}\selectfont
    \renewcommand{\arraystretch}{0.9}
    \setlength{\tabcolsep}{3pt}
    \robustify\bfseries
\robustify\itshape
\robustify\uline
\sisetup{uncertainty-mode=separate, round-mode=uncertainty, round-precision=1, table-align-uncertainty=true, mode=text, detect-mode=true, detect-weight=true, detect-shape=true}
\begin{tabular}{@{}lS[table-format = 1.3(1)] S[table-format = 1.3(1)] S[table-format = 1.3(1)] r@{}} 
    \toprule 
    & \multicolumn{1}{c}{{Lung}} & \multicolumn{1}{c}{{Liver}} & \multicolumn{1}{c}{{Brain}} & Rank \\ 
    \midrule 
LD & 0.7       & 0.75      & 0.71      & 9\textsuperscript{ }\\ 
CNN-10 (2017) & \bfseries 0.7995 +- 0.0087 & \bfseries 0.8822 +- 0.0219 & \bfseries 0.94 +- 0.0218 & 4\textsuperscript{\textdagger}\\ 
RED-CNN (2017) & \itshape 0.7573 +- 0.0228 & \itshape 0.8028 +- 0.0425 & 0.9509 +- 0.0191 & 6\textsubscript{ }\\ 
WGAN-VGG (2017) & \bfseries 0.9808 +- 0.0096 & \bfseries 0.923 +- 0.0488 & \itshape 0.8611 +- 0.0674 & 4\textsuperscript{\textdagger}\\ 
ResNet (2018) & \itshape 0.7515 +- 0.0227 & \itshape 0.7906 +- 0.0609 & 0.9118 +- 0.0464 & 7\textsubscript{ } \\ 
QAE (2019) & \itshape 0.8326 +- 0.0222 & \bfseries 0.9577 +- 0.011 & 0.9504 +- 0.0236 & 2\textsubscript{ }\\ 
DU-GAN (2021) & \itshape 0.9646 +- 0.0066 & \bfseries 0.9666 +- 0.0082 & 0.9393 +- 0.0758 & 1\textsubscript{ }\\ 
TransCT (2021) & \itshape 0.8338 +- 0.0128 & \itshape 0.923 +- 0.0138 & \itshape 0.8816 +- 0.0427 & 3\textsubscript{ }\\ 
Bilateral (2022) & \itshape 0.6362 +- 0.0149 & \itshape 0.873 +- 0.0229 & \itshape 0.8726 +- 0.0057 & 8\textsubscript{ }\\ 
\bottomrule 
\end{tabular} 

    \vskip-\baselineskip
\end{table}
\begin{figure}[tb]
  \centering
  \includegraphics[width=\columnwidth]{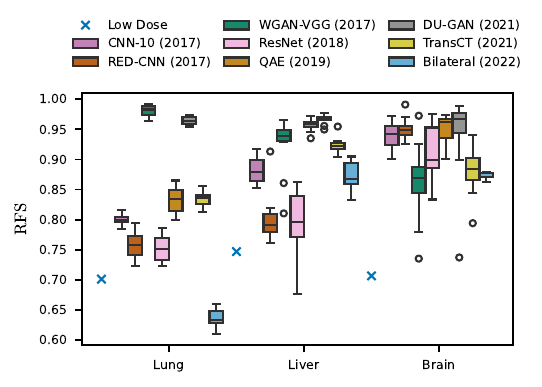}
  \vspace{-1.5\baselineskip}
  \caption{Radiomic feature similarity for different exam types and methods. Individual samples correspond to mean RFS over all scans of an anatomy for a single trained network. Box plots were then drawn over the ten training runs with different random seeds (\cf \cref{sec:benchmark_hpopt}).}
  \label{fig:rad_sim}
\end{figure}
We further evaluate all algorithms using the radiomic feature similarity in order to better assess whether the differences observed in the previous section translate to clinical features.

In \cref{fig:segmentations} we show contour plots of the automatic segmentations of the brain, lung, and liver for three high-dose scans of the test set. We visually verify that segmentations are reasonably good for all 15 scans in the test set. Those segmentation masks are then used to extract radiomic features for all low- and high-dose, as well as all denoised volumes of the test set. Using the same segmentation mask for subsequent radiomic feature extraction of all algorithms ensures a fair comparison, despite possible small errors produced by the automatic segmentation pipeline.

Upon evaluation of the radiomic feature similarity (\cref{tab:results_radsim} \& \cref{fig:rad_sim}), we find that radiomic features extracted for all denoising methods are significantly more similar to those extracted on the high-dose scan, compared to features extracted on the low dose scan, with Bilateral on lung data being the only exception. We also find that contrary to our findings using standard image quality metrics, RED-CNN is outperformed by numerous other algorithms, including the (older) CNN-10, and newer algorithms such as WGAN-VGG and QAE. Remarkably, the two GAN-based algorithms WGAN-VGG and DUGAN outperform all other algorithms on the lung data by a large margin. We hypothesize that this is due to the lower dose (10\% vs. 25\% for all other anatomies) on that data and the ability of GANs to produce more realistic noise textures in high-ambiguity settings compared to methods trained with standard pixelwise loss functions \cite{zhao2017}. Nonetheless, we do not find newer algorithms to consistently outperform older ones, and particularly the two newest algorithms considered in our study (TransCT and Bilateral) perform significantly worse \wrt radiomic feature similarity of all organs compared to older methods.
\subsection{Evaluation on LDCT-hard datasets}
\begin{figure}[tb]
    \centering
    \includegraphics[width=\columnwidth]{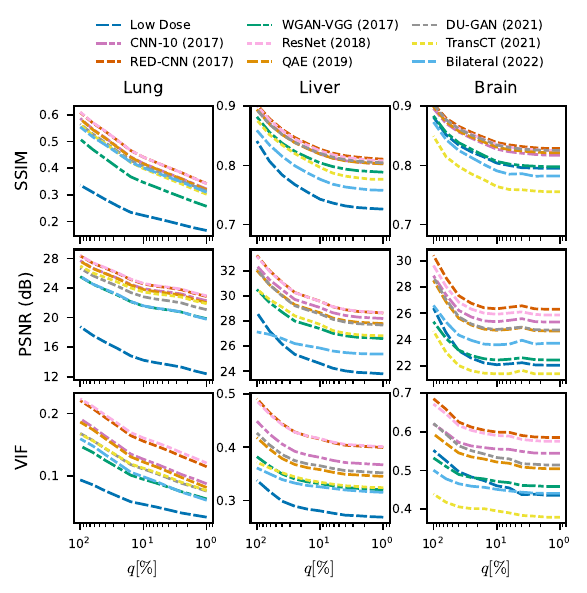}
    \vspace{-1.5\baselineskip}
    \caption{Evaluation of all methods for LDCT-hard-q\% for different values of q (right is smaller). For some settings and anatomies, methods perform up to 50\% worse for small q. The regular test set corresponds to q=100\%. Errorbars were omitted to improve visibility.}
    \label{fig:ldct_hard_curves}
    \vspace{-1\baselineskip}
\end{figure}
\looseness -1 \Cref{fig:ldct_hard_curves} shows the performance of individual methods for increasingly hard subsets of the training data (\ie, smaller $q$). We find a strong correlation between metrics for each method and the low dose scan. Although not surprising, this indicates that methods perform increasingly worse for increasing deviations of the low dose scan from the high dose scan. Additionally, the ranking among methods remains mostly invariant to $q$, and thus we conclude that all methods are similar in terms of their robustness to different amounts of deterioration of the low dose scan. Remarkably, WGAN-VGG, having a lower VIF and PSNR compared to the low dose scan on head exams for the regular test set (corresponding to $q=100\%)$, has a higher VIF and PSNR compared to the low dose scan for more difficult slices ($q\geq 16\%$ for VIF, $q\geq 40\%$ for PSNR). This may be explained by the aforementioned ability of GANs to produce more realistic results in high-ambiguity settings compared to networks trained in a pixelwise fashion. 

\Cref{fig:hardest_easiest} shows qualitative results for the slices from the test dataset for which the average SSIM over all methods is lowest (-) and highest (+), respectively. As can be seen, for each anatomy, the slice maximizing the average SSIM is the one where the cross-sectional area of the patient is small, thus reducing the noise in the low dose image.

\section{Discussion}
\label{sec:discussion}
In this study, we revisited some of the numerous proposed deep learning-based algorithms for low dose CT image denoising. We discovered several limitations in the experimental setups of these methods that hinder the verifiability of their claimed improvements. To overcome these challenges, we proposed a novel benchmark setup that promotes fair and reproducible evaluations. The setup comprises a unified data preprocessing, rigorous hyperparameter optimization, and evaluation using various metrics, including a novel metric that measures the similarity of radiomic features between the denoised volume and the high dose scan.

Upon evaluation of eight deep-learning based denoising algorithms proposed over the past six years, we find that there has been little progress. Particularly, when evaluated using standard image quality measures such as SSIM and PSNR, we find that no method consistently outperforms one of the earliest methods, RED-CNN. When evaluated using the radiomic feature similarity, we find that algorithms trained with an adversarial loss significantly outperform methods trained with pixel-wise losses on some data, indicating that the radiomic feature similarity provides useful information beyond standard, nonclinical image quality metrics. Nonetheless, the newest algorithms considered in our study fail to consistently outperform older ones. We also evaluated all methods on subsets of the test data consisting of increasingly difficult slices and find that methods are similarly robust to different amounts of deterioration of the low dose scan.

Similar to `reality checks' in related fields \cite{melis2018, musgrave2020}, our study highlights the need for a more rigorous and fair evaluation of novel deep learning-based denoising methods for low dose CT image denoising. We believe that our benchmark setup is a first and important step towards this direction and will help to develop novel and better algorithms.

\bibliographystyle{IEEEtran}
\bibliography{ldct-bench,MK}

\end{document}